\documentclass[aps,pre,twocolumn,superscriptaddress]{revtex4-1}
\usepackage{graphicx}
\usepackage{dcolumn}
\usepackage{bm}
\usepackage{amsmath}
\usepackage{amssymb}

\begin{document}

\title{Spatiotemporal patterns in the active cyclic Potts model}

\author{Hiroshi Noguchi}
\email[]{noguchi@issp.u-tokyo.ac.jp}
\affiliation{Institute for Solid State Physics, University of Tokyo, Kashiwa, Chiba 277-8581, Japan}
\author{Jean-Baptiste Fournier}
\email[]{jean-baptiste.fournier@u-paris.fr}
\affiliation{Laboratoire Mati\`ere et Syst\`emes Complexes (MSC), Universit\'e Paris Cit\'e \& CNRS, 75013 Paris, France}

\begin{abstract}
The nonequilibrium dynamics of a cycling three-state Potts model is studied on a square lattice using Monte Carlo simulations and continuum theory.
This model is relevant to chemical reactions on a catalytic surface and to molecular transport across a membrane. Several characteristic modes are formed depending on the flipping energies between successive states and the contact energies between neighboring sites. Under cyclic symmetry conditions, cycling homogeneous phases and spiral waves
form at low and high flipping energies, respectively. In the intermediate flipping energy regime, these two modes coexist temporally in small systems and/or at low contact energies. Under asymmetric conditions, we observed small biphasic domains exhibiting amoeba-like locomotion and temporal coexistence of spiral waves and a dominant non-cyclic one-state phase. An increase in the flipping energy between two successive states, say state 0 and state 1, while keeping the other flipping energies constant, induces the formation of the third phase (state 2), 
owing to the suppression of the nucleation of state 0 domains. Under asymmetric conditions regarding the contact energies, two different modes can appear depending on the initial state, due to a hysteresis phenomenon.
\end{abstract}

\maketitle

\section{Introduction}\label{sec:intro}

Spatiotemporal patterns are widely observed in nonequilibrium conditions, like waves of target and spiral shapes in two-dimensional (2D) systems~\cite{murr03,mikh06,mikh09,beta17,bail22,nogu24c,kura84,okuz03,qu06,sugi15}.
Cyclic dynamics of three or more states are one of the conditions to generate such waves.
The cyclic states are observed in various systems, such as chemical reactions,
gene expression, and ecological systems~\cite{mikh06,mikh09,kerr02,kels15}.
These dynamics are often explained using deterministic continuum equations~\cite{mikh06,mikh09,szol14,reic07,reic08} and lattice Lotka--Volterra models (also called rock--paper--scissors models)~\cite{kerr02,kels15,szol14,reic07,reic08,itoh94,tain94,szab99,szab02,john02,szcz13,juul13,mir22}.

The spatiotemporal patterns under noise and fluctuations are not yet fully understood.
Noise is typically used to trigger an excitable wave
and to generate stochastic resonance~\cite{gamm98,mcdo09}.
The wave propagation of subexitable media of photosensitive Belousov--Zhabotinsky reaction 
is enhanced by the random variation of light intensity~\cite{mikh06,kada98,wang99,alon01}.
In partial differential equations, random noises are added to include mesoscale fluctuations~\cite{hild96,reic07,reic08}.
In the lattice Lotka--Volterra models~\cite{kerr02,kels15,szol14,reic07,reic08,itoh94,tain94,szab99,szab02,john02,szcz13,juul13,mir22} for the predator--prey systems,
noise is included as a random selection of lattice sites.
Since populations increase by self-reproduction, the nucleation of species domains is not considered.
Hence, the extinct species do not reapear (i.e., absorbing transition).
In contrast, in small molecular systems, thermal fluctuations have dominant effects,
and nucleation and growth are the crucial kinetic processes.
Spiral and target wave patterns have been observed in chemical reactions on noble metal surfaces~\cite{ertl08,bar94,goro94,barr20,zein22}.
In a previous paper~\cite{nogu24}, we studied a nonequilibrium three-state Potts model
under cyclic symmetry (the three states being equivalent), and reported that nucleation and growth can alter spatiotemporal patterns.
We found two modes: 
(i) a homogeneous cycling mode (HC), where each state dominantly covers the system while changing cyclically through nucleation and growth.
(ii) a spiral wave (SW) mode where the three states coexist while spiral waves are formed from the contact points of three states.
These two modes can temporally coexist in small systems but not in larger systems.
In this study, we examine the dynamics of the three-state active Potts model in asymmetric cycling conditions.
We will show several modes including hysteresis and amoeba-like motions of small biphasic domains. A non-cyclic one-state dominant phase and temporal coexistence with the SW mode newly appear.

The cyclic Potts model and method are described in Sec.~\ref{sec:model}.
It is a 2D lattice model for chemical reactions on a catalytic surface and molecular transport through a membrane.
The dynamics under cyclic-symmetry conditions are briefly described in Sec.~\ref{sec:sym}.
The dynamics in asymmetric flipping energy and asymmetric contact energies
are described in Secs.~\ref{sec:asymf} and \ref{sec:asyme}, respectively.
The theoretical analysis by a continuum theory is described in Sec.~\ref{sec:theory}.
Finally, a summary is presented in Sec.~\ref{sec:sum}.

\begin{figure*}[tbh]
\includegraphics[]{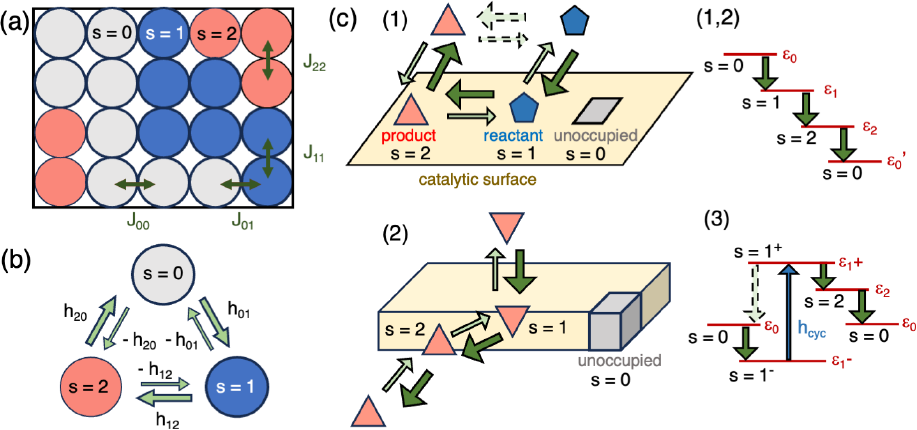}
\caption{
(a) Three-state Potts model in a square lattice.
(b) Nonequilibrium cyclic flipping of the three states driven by $h_{01}+h_{12}+h_{20}=h_{\mathrm{cyc}}\ne 0$. 
(c) Possible realizations with energy levels depicted: (1) Chemical reaction on a catalytic surface, (2) Molecular transport through a membrane, (3) Three-state process with $1^-\to1^+$ excitation.
The wide and narrow arrows represent the dominant (forward) and secondary (backward) processes, respectively. The dashed arrows represent kinetically frozen processes.
}
\label{fig:cart}
\end{figure*}

\section{Active Cyclic Potts Model}\label{sec:model}

We consider a 2D square lattice with sites $i$ having three states $s_i=\alpha$, with $\alpha\in\{0,1,2\}$,
as shown in Fig.~\ref{fig:cart}(a).
Two nearest neighbor sites have a contact energy $-J_{s_is_j}$.
In addition, the three states have different self-energies $\varepsilon_\alpha$, so that the total Hamiltonian reads
\begin{equation}
\label{StartingHamiltonian}
H = H_{\mathrm{int}} + \sum_i\varepsilon_{s_i},\quad H_{\mathrm{int}}= - \sum_{\langle ij\rangle} J_{s_is_j}.
\end{equation}
To model the attraction between like states, we mainly use symmetric contact energies $J_{\alpha\alpha}=J$, and set $J_{\alpha\beta}=0$ for $\alpha \ne \beta$.
This corresponds to the standard three-state Potts model with external fields~\cite{pott52,bind80}.
We define the equilibrium flipping energies $h_{\alpha\beta}$ as the variations $h_{\alpha\beta}=\varepsilon_\alpha-\varepsilon_\beta$. 
Hence, $h_{01}+h_{12}+h_{20}=0$ by definition in thermal equilibrium conditions.

We extended this model to the nonequilibrium situation in which $h_{\alpha\beta}\ne\varepsilon_\alpha-\varepsilon_\beta$, but where still $h_{\beta\alpha}=-h_{\alpha\beta}$. Then we have $h_{01}+h_{12}+h_{20}= h_{\mathrm{cyc}} \ne 0$ (see Fig.~\ref{fig:cart}(b)). A possible way to do this is to add a nonequilibrium or active contribution to some or all of the flipping energies, in the form $h_{\alpha\beta}=\varepsilon_\alpha-\varepsilon_\beta+h^{\mathrm{neq}}_{\alpha\beta}$.
Hence, the detailed-balance condition is locally satisfied for one flip between states $s=\alpha$ and $\beta$ but not globally for cycles ($s=0 \to 1 \to 2 \to 0$).

Three possible designs are described below: chemical reaction, molecular transport, and excitation, as shown in Fig.~\ref{fig:cart}(c).
(1) Reaction on a catalytic surface: 
molecules bind and unbind to the surface, with the state $s=0$ corresponding to an empty surface site, and the other two states to a site occupied by a molecule, either in the form $s=1$ or $s=2$.  
The surface catalyzes the reaction from $s=1$ to $s=2$, whereas this reaction is kinetically frozen in the bulk. 
(2) Molecular transport between two sides of a membrane: amphiphilic molecules bind to both sides of a bilayer membrane (states $s=1$ and $2$) and switch between them (flip--flop).
The molecules are transported by a difference in chemical potential between the solutions on the two sides of the membrane~\cite{miel20,holl21,nogu23}.
(3)  An excitation process: $s=1$  has a low-energy state $s=1^-$ and a high-energy state $s=1^+$ triggered by external means (e.g., photoexcitation or ATP hydrolysis). Then, if the backward transformation from $s=1^+$ to $s=0$ is negligibly slow, we have effectively $h_{\mathrm{cyc}}=\varepsilon_{1^+}-\varepsilon_{1^-}>0$. 
Experimentally, chemical waves have been observed at catalytic surfaces (H$_2$ or CO oxidation and NO or NO$_2$ reduction on noble metal surfaces such as palladium and ruthenium)~\cite{ertl08,bar94,goro94,barr20,zein22}.
The transfer of water molecules through a chiral liquid-crystalline monolayer induces a target wave~\cite{tabe03}.
Although multiple reactions or complicated molecular interactions occur in these experimental systems, 
we constructed a simple minimum model to capture the essence.

For simplicity, no state exchange between neighboring sites (diffusion) is considered.
The thermal energy and the lattice spacing are normalized to unity.
We mainly use $J_{00}=J_{11}=J_{22}=J=2$ to induce a phase separation between different states when using symmetric conditions for the contact energy.
To examine the effects of asymmetric contact energies, $J_{00}$ is varied from $1.75$ to $2.8$  with keeping $J_{11}=J_{22}=2$, in Sec.~\ref{sec:asyme}.
Our lattice has $N$ sites with a side length of $\sqrt{N}$ under periodic boundary conditions.
The state of a site is flipped according to a Monte Carlo (MC) method.
A site is chosen at random, and then it is flipped to either of the other two states with $1/2$ probability.
The new state is accepted or rejected with Metropolis probability
\begin{equation}\label{eq:mc}
p_{s_is'_i}=\min\left(1,e^{-\Delta H_{s_is'_i}}\right)
\end{equation}
in which $\Delta H_{s_is'_i}=H'_{\mathrm {int}}-H_{\mathrm{int}}-h_{s_is'_i}$ is the energy variation in the change from the old state to the new one. This procedure is performed $N$ times per MC step (time unit). Statistical errors are calculated from three or more independent runs.

\begin{figure*}[tbh]
\includegraphics[]{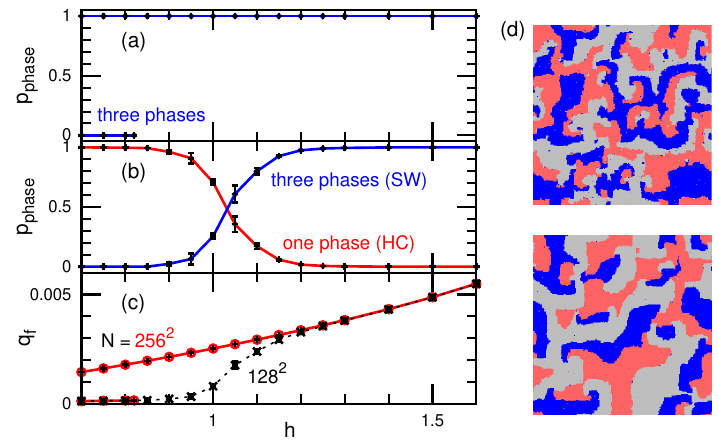}
\caption{
Cyclic flipping in symmetric condition ($h_{01}=h_{12}=h_{20}=h$ and $J_{\alpha\alpha}=2$).
(a) System of size $N=256^2$. Probability $p_{\mathrm{phase}}=p_{\mathrm{three}}$ of the three-phase state ($N_0>0.02N$, $N_1>0.02N$, and $N_2>0.02N$). At $h \leq 0.82$, depending on the initial condition, both the SW and the HC modes (indicated by the vanishing of $p_{\mathrm{three}}$) can appear.
At $h \geq 0.83$, only the SW mode can exist.
(b) System of size $N=128^2$. Probability $p_{\mathrm{three}}$ of the three-phase state and probability $p_{\mathrm{one}}$ of one-phase state ($N_s>0.98N$ with $s\in \{0,1,2\}$).
The HC and SW modes temporally coexist for $1 \leq h \leq 1.15$.
(c) Average difference flow $q_{\mathrm{f}}$ between forward and backward flips, showing that detailed balance is more violated in the SW mode than in the HC mode.
The solid and dashed lines represent the data at $N=256^2$ and $128^2$, respectively.
(d) Snapshots at $h=0.9$ and $1.6$ from bottom to top for $N=256^2$.
The light gray, blue (dark gray), and red (medium gray) represent $s=0$, $1$, and $2$, respectively.
}
\label{fig:psym}
\end{figure*}

\begin{figure*}[tbh]
\includegraphics[]{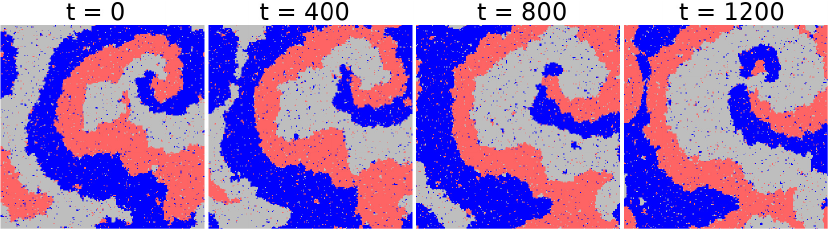}
\caption{
Sequential snapshots at $h_{01}=h_{12}=h_{20}=h=0.25$ and $J_{\alpha\alpha}=1.2$, in symmetric condition at $N=256^2$.
}
\label{fig:e12a}
\end{figure*}

\begin{figure*}[tbh]
\includegraphics[]{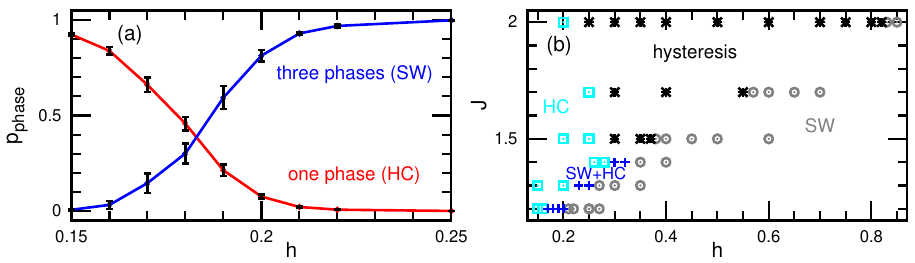}
\caption{Dependence of the system state on the contact energy $J$ and the flipping energy $h$ in symmetric condition ($h_{01}=h_{12}=h_{20}=h$) at $N=256^2$.
(a) Probabilities $p_\mathrm{phase}=p_{\mathrm{three}}$ (blue) of the three-phase state and $p_\mathrm{phase}=p_{\mathrm{one}}$ (red) of the one-phase state ($N_s>0.95N$ with $s\in \{0,1,2\}$) at $J_{\alpha\alpha}=1.2$ as a function of $h$.
The HC and SW modes temporally coexist for $0.17 \leq h \leq 0.2$.
(b) Dynamic phase diagram (for an observation time $t\sim10^8$).
The squares (light blue) and circles (gray) represent the HC and SW modes, respectively.
The pluses (blue) represent the temporal coexistence of the SW and HC modes (SW+HC).
The asterisks (black) represent the hysteresis region of the SW and HC modes. Either SW or HC mode appears depending on initial states.
}
\label{fig:e12b}
\end{figure*}

\section{Dynamics in Cyclic-Symmetry Conditions}\label{sec:sym}

First, we briefly describe the dynamics of the active cyclic Potts model in symmetric conditions, i.e., for 
$h_{01}=h_{12}=h_{20}=h$ and $J_{00}=J_{11}=J_{22}=J$. 
Detailed results at $J=2$ are given in Ref.~\cite{nogu24}.
For large systems ($N\geq 192^2$), at low $h$, either the HC or SW mode appears (see their definitions in Sec.~\ref{sec:intro}), depending on the initial state,
while only the SW mode appears at high $h$.
Transitions from the HC to SW modes were observed, but the reverse transition never occurred (within accessible simulation times).
In the SW mode, spiral waves propagate from the contact points of three states,
and the number of each state fluctuates around $N/3$ (see the snapshots in Fig.~\ref{fig:psym}(d)).
For small systems ($N\leq 154^2$), 
the HC mode appears at low $h$, the SW mode appears at high $h$, 
and the two modes temporally coexist at medium $h$.
Representing temporal coexistence by a `+', we denote this mixed state by SW+HC.

The ratios of these two modes can be quantified using the number densities $N_\alpha/N$ of the three states.
For $N_\alpha>0.98N$, we consider that the lattice is dominantly covered by a unique state $s=\alpha$, and we call this global state `one' phase, unless otherwise specified.
The fraction of time spent in this state is denoted by $p_{\mathrm{phase}}=p_{\mathrm{one}}$ (Fig.~\ref{fig:psym}).
In the HC mode, this `one' phase global state cyclically changes from $s=\alpha$ to $[\alpha+1]$, where $[\alpha+1]=(\alpha+1)\bmod 3$.
Since the transient dynamics of nucleation and growth is rapid, we identify $p_{\mathrm{one}}$ with the fraction of time spent in the HC mode in the cyclic-symmetry condition.
Moreover, we consider that the system is in the `three' phases global state when $N_0>0.02N$, $N_1>0.02N$, and $N_2>0.02N$. The fraction of time spent in this state is called $p_{\mathrm{phase}}=p_{\mathrm{three}}$ (Fig.~\ref{fig:psym}).
The  ratio $p_{\mathrm{one}}/p_{\mathrm{three}}$ corresponds to the ratio of the times spent in the HC and SW modes in the cyclic-symmetry condition
(see  Figs.~\ref{fig:psym}(a) and (b)).

Since the nucleation barrier decreases with increasing $h$ (at fixed contact energy $J$) and the nucleation more frequently occurs in larger systems, the mean lifetime of the `one phase' state was found to be roughly proportional to $\exp(-h)/N$~\cite{nogu24}.
Furthermore, the domains were found to grow with a velocity of $v_{\mathrm {wave}} \simeq 0.07h + 0.009$ at $h \ge 0.5$.
Hence, at high $h$ and/or large $N$, during the growth of a $s=[\alpha+1]$ domain within a $s=\alpha$ domain, the nucleation of a smaller $s=[\alpha+2]$ domain often starts. Because domain growth is a stochastic process, three-state contacts are thus frequently produced.
When a three-state contact appears, a spiral wave forms, as the three domain boundaries associated with this contact point exhibit a rotative motion (each $s=\alpha$ phase is invaded by the adjacent $s=[\alpha+1]$ phase).
Conversely, the SW mode changes into the HC mode, when the three-state contact points stochastically disappear.
This disappearance occurs in small systems but not in large systems ($N\geq 192^2$) on the simulation time scale ($t \sim 10^8$--$10^9$).
Similar size dependence of spiral waves was reported in excitable media~\cite{qu06,sugi15}.
Note that the continuous transition via the temporal coexistence likely recovers in the long-time limit ($t\to \infty$) even for large systems.
Since the density of the three-state contacts linearly increases with increasing $h$ (compare two snapshots in Fig.~\ref{fig:psym}(d)),
the SW mode more often changes into the HC mode at lower $h$. With increasing system size $N$, the transition between the HC and SW modes occurs at lower $h$, and the coexistence region becomes narrower. In the large-size limit ($N\to \infty$), the HC mode does not likely appear for finite $h$ and $t\to \infty$ (the transition point $h \to 0$).
More detail is given in Ref.~\cite{nogu24}.

During the cyclic flipping,
the forward flip ($s=\alpha\to [\alpha+1]$) more frequently occurs than the backward flip ($s=[\alpha+1]\to \alpha$).
This is quantified by the flow $q_\mathrm{f}$ defined as the average difference between forward and backward flips per site, during one MC step.
Figure \ref{fig:psym}(c) shows that $q_\mathrm{f}$ is much higher in the SW mode than in the HC mode
and increases with increasing $h$ in both modes.

As the contact energy $J$ decreases, the SW mode appears at lower $h$, and the cyclic change of dominant phases in the HC mode becomes faster. Thus, the temporal coexistence of these two modes can be observed in large systems. For $N=256^2$, the coexistence is obtained at $J=1.2$, as shown in Figs.~\ref{fig:e12a} and \ref{fig:e12b}(a). Note that the threshold $N_\alpha/N= 0.95$ is used for the one-phase state since the domains contain 3\% of the other states at $J=1.2$ (see Fig.~\ref{fig:e12a}).
The transition between the HC and SW modes occurs at higher $h$ with increasing $J$.
For $J\gtrsim 1.5$, the discontinuous transition is obtained and the two modes maintain for $t \sim 10^8$ around the transition point owing to hysteresis (see Fig.~\ref{fig:e12b}(b)). However, this transition likely becomes continuous on a longer time scale.
For $J=2$, the SW mode changes into the one-phase dominant phase at low $h$ ($h \lesssim 0.2$).
This phase cycles as the HC mode in the long-time limit but does not cycle to the next phase on the simulation time scale,
since the nucleation duration becomes exponentially long with decreasing $h$~\cite{nogu24}.

\begin{figure*}[tbh]
\includegraphics[]{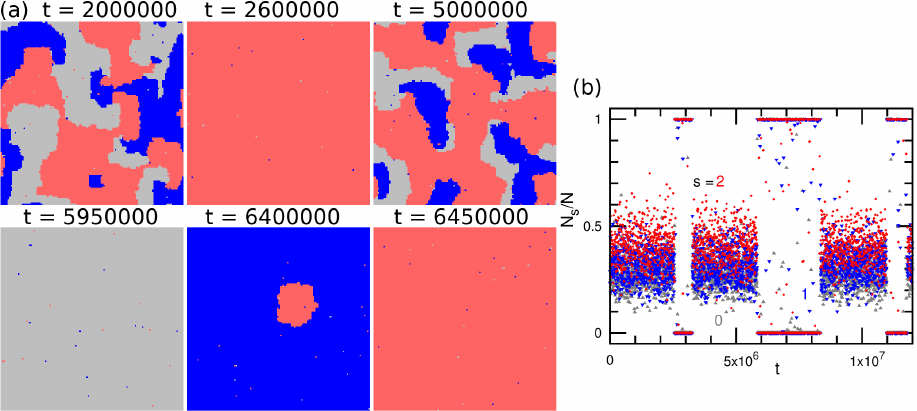}
\caption{
Behavior of the system for weakly asymmetric flipping energies. Temporal coexistence of the SW and HC modes (SW+HC) for $h_{01}=1.1$, $h_{12}=h_{20}=1$,  $J_{\alpha\alpha}=2$, and $N=128^2$.
(a) Sequential snapshots.
(b) Time evolution of the fraction of sites in each state (corresponding to snapshots in (a)).
The light gray, blue (dark gray), and red (medium gray) colors represent the data for $s=0$, $1$, and $2$, respectively.
}
\label{fig:tm1d1n11}
\end{figure*}

\begin{figure*}[tbh]
\includegraphics[]{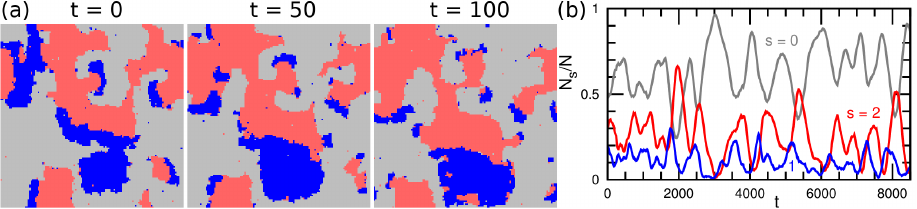}
\caption{
Behavior of the SW mode in the case of a large asymmetry between the flipping energies, for $h_{01}=1.6$, $h_{12}=1.7$, $h_{20}=1$,  $J_{\alpha\alpha}=2$, and $N=128^2$.
(a) Sequential snapshots.
The interfaces between the $s=1$ and $s=2$ domains, which move towards the $s=1$ domain, are faster than the other boundaries.
(b) Time evolution of the fractions of sites in each state (corresponding to snapshots in (a)).
The light gray, blue (dark gray), and red (medium gray) colors represent the data for $s=0$, $1$, and $2$, respectively.
}
\label{fig:tm1d17n16}
\end{figure*}

\begin{figure*}[tbh]
\includegraphics[]{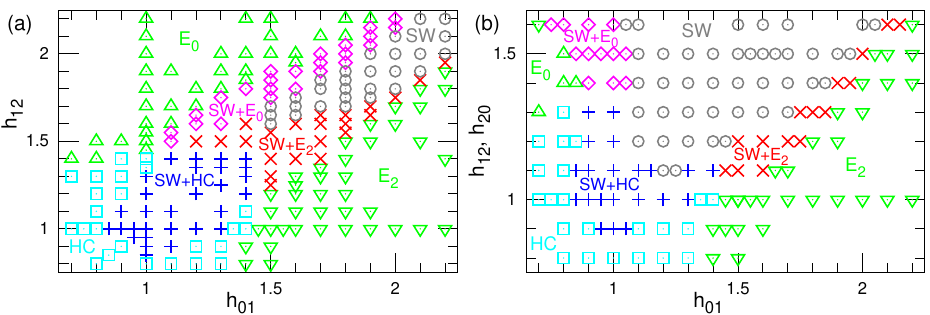}
\caption{
Dynamic phase diagrams in the case of asymmetric flipping energies for $J_{\alpha\alpha}=2$ and $N=128^2$.
(a) Diagram as a function of $h_{01}$ and $h_{12}$ for $h_{20}=1$.
(b) Diagram as a function $h_{01}$ and $h_{12}=h_{20}$.
The squares (light blue) and circles (gray) represent the HC and SW modes, respectively.
The up-pointing and down-pointing triangles (green) represent the homogeneous phases $s=0$ (denoted E$_0$) and $s=2$ (denoted E$_2$),
respectively.
The pluses (blue), diamonds (magenta), and crosses (red) represent the temporal coexistence
SW+HC, SW+E$_0$, and SW+E$_2$, respectively.
The diagonal in (b) corresponds to the cyclic-symmetry condition ($h_{01}=h_{12}=h_{20}$).
}
\label{fig:pdh}
\end{figure*}

\begin{figure*}[tbh]
\includegraphics[]{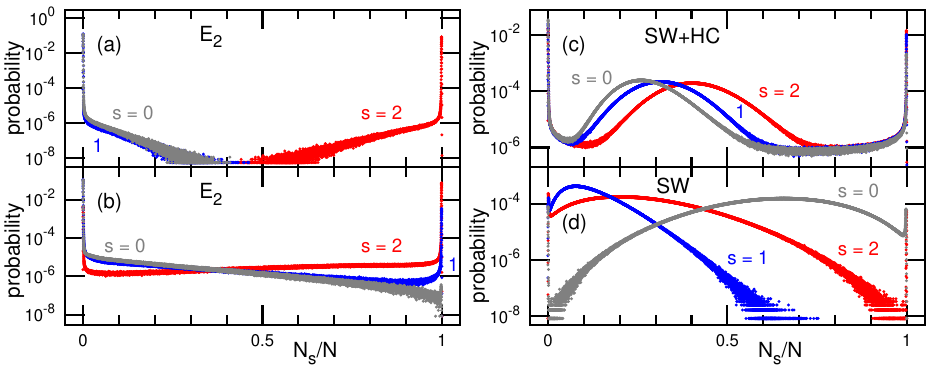}
\caption{
Probabilities to observe given fractions $N_s/N$ of each state in the case of asymmetric flipping energies for $J_{\alpha\alpha}=2$ and $N=128^2$.
(a) State E$_2$ (homogeneous $s=2$ phase) for $h_{01}=1.7$ and $h_{12}=h_{20}=1$. (b) State E$_2$ (close to the HC mode) for $h_{01}=1.45$ and $h_{12}=h_{20}=1$.
(c) Temporal coexistence of the SW and HC modes (SW+HC) for $h_{01}=1.1$ and $h_{12}=h_{20}=1$, corresponding to Fig.~\ref{fig:tm1d1n11}.
(d) SW mode at $h_{01}=1.6$, $h_{12}=1.7$, and $h_{20}=1$, corresponding to Fig.~\ref{fig:tm1d17n16}.
}
\label{fig:nhis}
\end{figure*}

\begin{figure}[tbh]
\includegraphics[]{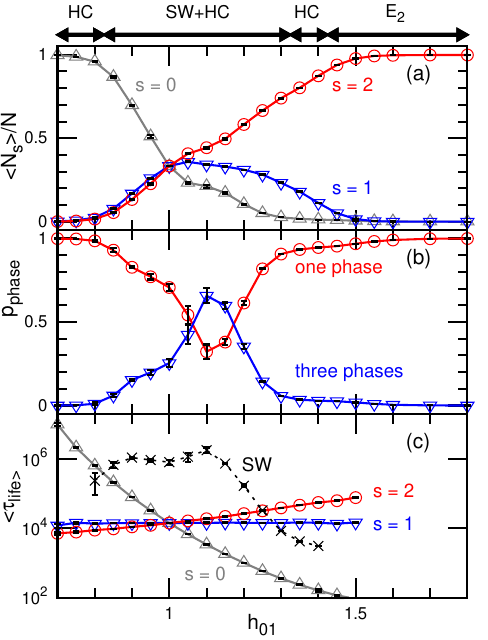}
\caption{
Density of states and mode lifetimes in the case of asymmetric flipping energies as a function of $h_{01}$ for $h_{12}=h_{20}=1$, $J_{\alpha\alpha}=2$, and $N=128^2$.
The dynamic mode in which the system sits is indicated at the top of the figure. (a) Fraction of sites in each state. The fully symmetric case corresponds to the crossing of the three lines at $h_{01}=1$. For $h_{01}<1$ the $s=0$ state dominates even in the HC mode, since the system spends more time in this state.
The gray up-pointing triangles, blue down-pointing triangles, and red circles represent the data for $s=0$, $1$, and $2$, respectively.
(b) Probabilities of one-phase state and three-phase coexistence.
(c) Mean lifetimes of the homogeneous phases (solid lines) and SW mode (dashed line).}
\label{fig:m1d1}
\end{figure}

\begin{figure}[tbh]
\includegraphics[]{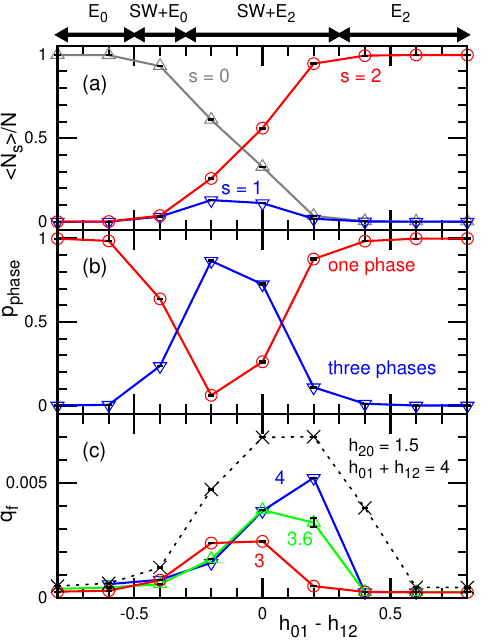}
\caption{
Dependence of the system's properties on the asymmetry $h_{01}-h_{12}$ when keeping $h_{01}+h_{12}$ constant, for $J_{\alpha\alpha}=2$ and $N=128^2$.
(a), (b) $h_{01}+h_{12}=3$  and  $h_{20}=1$.
(a) Fraction of sites in each state.
The gray up-pointing triangles, blue down-pointing triangles, and red circles represent the data for $s=0$, $1$, and $2$, respectively.
(b) Probabilities of the one-phase state and the three-phase coexistence.
The corresponding dynamic modes are described at the top of the figure.
(c) Average difference flow $q_{\mathrm{f}}$ between forward and backward flips.
The solid lines represent the data for $h_{01}+h_{12}=3$, $3.6$, and $4$ at $h_{20}=1$.
The dashed lines represent the data for $h_{01}+h_{12}=4$ and $h_{20}=1.5$.
}
\label{fig:mdsa}
\end{figure}

\begin{figure*}[tbh]
\includegraphics[]{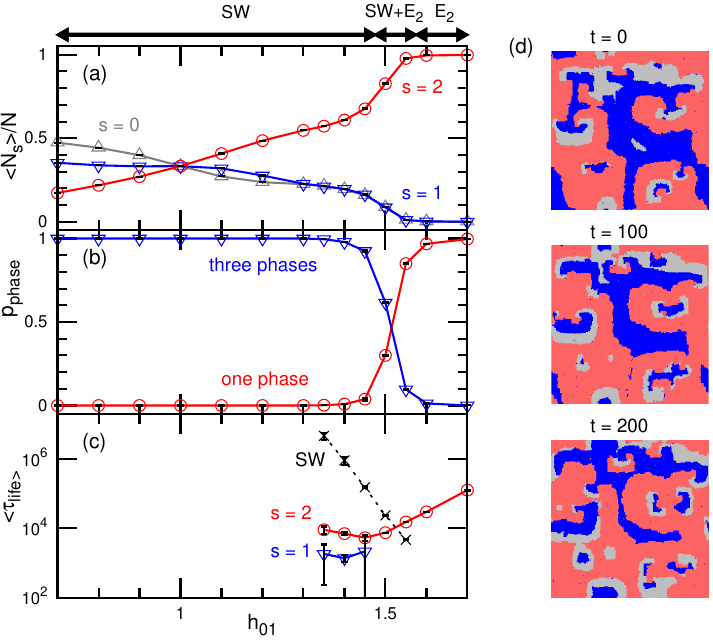}
\caption{
Density of states and mode lifetimes in the case of asymmetric flipping energies as a function of $h_{01}$ for $h_{12}=h_{20}=1$, $J_{\alpha\alpha}=2$, and $N=256^2$.
(a) Fraction of sites in each state.
The gray up-pointing triangles, blue down-pointing triangles, and red circles represent the data for $s=0$, $1$, and $2$, respectively.
(b) Probabilities of one-phase state and three-phase coexistence.with
(c) Mean lifetimes of the homogeneous phases (solid lines) and SW mode (dashed line).
The dynamic modes are described at the top of the figure.
(d) Sequential snapshots at $h_{01}=1.3$.
}
\label{fig:n256}
\end{figure*}

\section{Dynamics for Asymmetric Flipping Energies}\label{sec:asymf}

In this section, we describe the dynamics with asymmetric flipping energies, for symmetric contact energies $J_{\alpha\alpha}=2$. By asymmetric flipping, we mean that $h_{01}$, $h_{12}$, and $h_{20}$ are not all equal. We still assume, however, that $h_{01}+h_{12}+h_{20}= h_{\mathrm{cyc}} \ne 0$ and 
keep $h_{\beta\alpha}=-h_{\alpha\beta}$.
First, we consider a small system of $N=128^2$ sites, in which the HC and SW modes can coexist in the symmetric condition.
When the flipping energies slightly deviate from the symmetric condition,
the dynamics exhibit only small changes.
For example, the SW+HC mode is obtained at $h_{01}=1.1$, $h_{12}=h_{20}=1$, similar to that at $h_{01}=h_{12}=h_{20}=1$, and the mean densities of $s=0$ and $s=2$ are slightly lower and higher than $1/3$ in the SW mode, respectively (see Fig.~\ref{fig:tm1d1n11} and Movies S1 and S2).

However, large asymmetry alters the dynamics qualitatively. 
In symmetric conditions, the three-phase contact points move isotropically,
and the boundaries between domains move all at the same speed in the direction from $s=[\alpha+1]$ to $s=\alpha$.
In the case of asymmetric flipping energies, the speeds of the different boundaries are not the same.
In the SW mode, when the flipping energies differ substantially,
the three-phase contact points move ballistically rather than diffusively (see Fig.~\ref{fig:tm1d17n16} and Movie S3).
For $h_{01}=1.6$, $h_{12}=1.7$, and $h_{20}=1$, as in Fig.~\ref{fig:tm1d17n16}, small biphasic domains of $s=1$ and $2$ move like amoeba locomotion and bacteria colony growth in the direction of the $s=1$ region. Their large fluctuations often result in domain division and disappearance.

In the asymmetric case, the average fraction $\rho_\alpha=N_\alpha/N$ of sites in each state differs from $1/3$, as can be seen in Fig.~\ref{fig:tm1d17n16}. The contact energies turn out to be essential in determining those fractions. Indeed, for $h_{01}=1.6$, $h_{12}=1.7$, and $h_{20}=1$, the mean-field approach developed in Sec.~\ref{sec:structureless}, which neglects both the contact energies and the spatial structures, predicts $\rho_0\simeq\rho_1\simeq0.32$ and $\rho_2\simeq0.37$, whereas for $J=2$ the actual ratio $\rho_0/\rho_1\approx5$ (see Fig.~\ref{fig:tm1d17n16}). This is expected, since a single site flip within a domain involves the loss of four $J_{\alpha\alpha}=J$ contacts, whereas a boundary flip involves the loss of two contacts ($2J=4$). The nucleation of a domain and the motion of an interface are, therefore, cooperative events, and they strongly affect the state ratios.

Dynamic phase diagrams are shown in Fig.~\ref{fig:pdh}.
In addition to the SW and HC modes, new one-phase modes E$_\alpha$, in which the state $s=\alpha$ is predominant as in equilibrium, appear when either $h_{01}$, $h_{12}$ or $h_{20}$ dominates.
Temporal coexistence SW+E$_\alpha$ is possible, as shown in Fig.~\ref{fig:pdh}.
The various modes are distinguished as follows.
When $p_{\mathrm{three}}>0.05$ (see its definition in Sec.~\ref{sec:sym}), the system is in the SW mode.
When $p_{\mathrm{one}}>0.05$, the system is either in the HC mode or in one of the E$_\alpha$ modes.
In the case both of them exceed $0.05$, the system is either in the coexistence mode SW+HC or in a coexistence mode SW+E$_\alpha$.
The HC and E$_\alpha$ modes are distinguished from the distribution of $N_s$ shown in Fig.~\ref{fig:nhis}.
When all three states have peaks at $N_s/N\approx 1$, the system is in the HC mode.
Otherwise, we consider it to be in the E$_{\alpha}$ mode, in which the $s=\alpha$ state has the maximum peak at $N_s/N\approx 1$.
Since a very small peak can be buried or formed by statistical error,
we consider that a peak is recognizably high as a dominant phase,
when the peak at $N_\alpha/N\approx 1$ is ten times higher than the local minimum close to $N_\alpha/N=1$
(see the lower peak of $s=0$ in Fig.~\ref{fig:nhis}(b), which is in the E$_2$ mode near the threshold to the HC mode).
Thus, the cycling of two homogeneous phases still continues near the mode boundary in the E$_\alpha$ region of the phase diagram, but the cycling completely disappears deeply in the E$_\alpha$ region (compare Fig.~\ref{fig:nhis}(a) and (b)).

As $h_{01}$ increases while $h_{12}$ and $h_{20}$ are fixed,
the dynamics changes from E$_0$, to SW+E$_0$, SW, SW+E$_2$, and  E$_2$ (or HC instead of E$_0$), in order (see Figs.~\ref{fig:pdh} and \ref{fig:m1d1}).
Interestingly, high $h_{01}$ generates the E$_2$ mode, although $h_{01}$ directly enhances the nucleation and growth of the $s=1$ phase.
This dynamics can be understood by seeing the lifetimes of one-state dominant phases shown in Fig.~\ref{fig:m1d1}(c).
The lifetime of $s=\alpha$ is calculated as the period during which the system stays in $N_{\alpha}>0.98N$.
The mean lifetime of the $s=0$ dominant phase exponentially decreases with increasing $h_{01}$,
by more frequent nucleation and growth of the $s=1$ phase in the $s=0$ phase, as expected.
Conversely, the mean lifetime of the $s=1$ phase is almost independent of $h_{01}$,
because the $s=1$ phase is terminated by the nucleation and growth of the $s=2$ phase, which is determined by $h_{12}$.
However, the $s=2$ phase remains for longer periods with increasing $h_{01}$,
owing to the suppression of the nucleation.
To understand this, let us consider an isolated dimer of $s=0$ sites in the $s=2$ phase.
One site in this dimer can go back to the $s=2$ state via two pathways.
One is the direct flipping to $s=2$, whose probability is min$(1,\exp(2J-h_{20}))$,
since the number of contacts between $s=0$ and $s=2$ sites changes from six to four.
The other is the two-step flipping via $s=1$, in which the probabilities of $s=0\to 1$ and $1\to 2$ 
are min$(1,\exp(-J+h_{01}))$ and min$(1,\exp(3J+h_{12}))$, respectively.
The latter flipping increases for higher $h_{01}$, since the first step is rate-limiting.
Thus, the $s=2$ phase becomes temporally dominant.
Conversely, the acceleration of the second step does not enhance the latter flipping, as seen in the constant lifetime of the $s=1$ state.

A similar tendency in the flipping energies is also found in the SW mode.
When $h_{\alpha[\alpha+1]}$ is the highest among the three flipping energies,
the $s=[\alpha+2]$ state is the most occupied over the lattice (see Figs.~\ref{fig:tm1d1n11} and \ref{fig:tm1d17n16}).
The lifetime of the SW mode is calculated as the lapse from the time entering $0.1N<N_\alpha<0.75N$ for all $\alpha$ to that entering the one phase mode ($N_\alpha>0.98N$ for one of the states).
When the three forward flips are comparable (e.g., $h_{01} \simeq 1$ while $h_{12}=h_{20}=1$ as in Fig.~\ref{fig:m1d1}),
the mean lifetime of the SW mode becomes longer and the system falls into this mode.
As the flipping energies deviate more from the symmetric condition,
one of the states ($s=\alpha$) dominates the SW mode,
and subsequently the SW mode changes into the $s=\alpha$ dominant phase, E$_\alpha$, via the SW+ E$_\alpha$ mode or the HC mode (see Fig.~\ref{fig:pdh}).

The phase diagram in Fig.~\ref{fig:pdh}(a) is slightly asymmetric with respect to the symmetric axis  $h_{01}=h_{12}$.
This is due to the fact that the transition from the $s=\alpha$  state to the $s=[\alpha+1]$ state is also dependent on the $[\alpha+1]\to [\alpha+2]$ flip, as mentioned above.
This asymmetry can be more quantitatively captured by plotting the various physical quantities as a function of $h_{01}-h_{12}$ along the diagonal line fixing $h_{01}+h_{12}$, as shown in Fig.~\ref{fig:mdsa}.
At $h_{01}+h_{12}=3$ and $h_{20}=1$,
the SW mode more often occurs for $h_{01}-h_{12}<0$ than for $h_{01}-h_{12}>0$, and
the center of the SW existence region is at  $h_{01}-h_{12} \simeq -0.1$ (see  Figs.~\ref{fig:mdsa}(a) and (b)).
The flow $q_{\mathrm{f}}$ has a maximum around this center (compare the red line in Fig.~\ref{fig:mdsa}(c) and the blue line in Fig.~\ref{fig:mdsa}(b)).
With increasing $h_{01}+h_{12}$, the center position is shifted to the positive direction
and the maximum of $q_{\mathrm{f}}$ becomes higher (see the three solid lines in Fig.~\ref{fig:mdsa}(c)).
At higher $h_{20}$, $q_{\mathrm{f}}$ increases (compare the top two lines in Fig.~\ref{fig:mdsa}(c)).
Thus, the asymmetry direction and amplitude vary depending on the conditions.

\begin{figure*}
\includegraphics[]{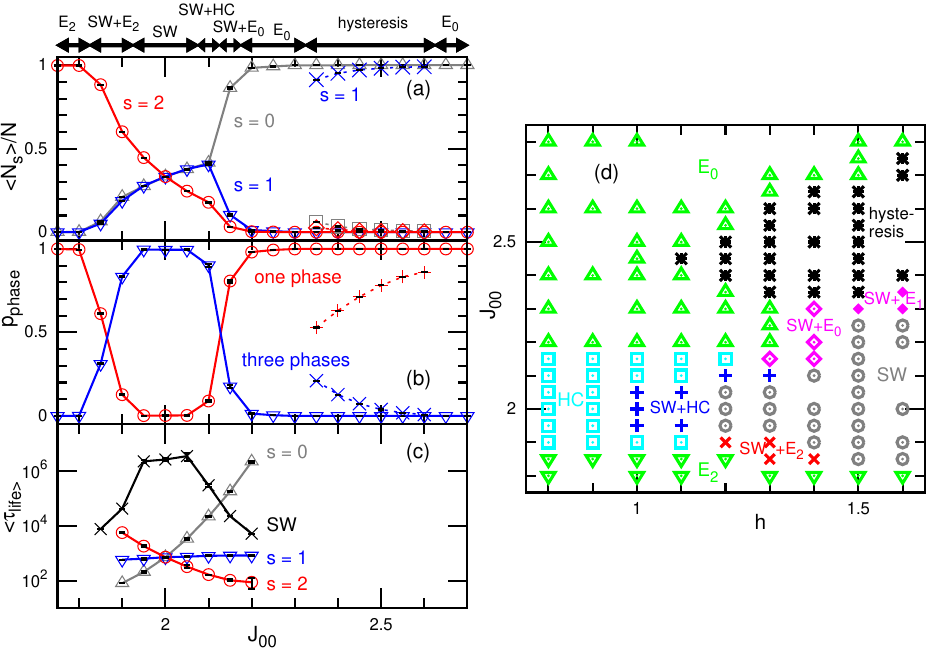}
\caption{
Effects of the asymmetric contact energies with the symmetric flipping energies $h=h_{01}=h_{12}=h_{20}$ at $J_{11}=J_{22}=2$ and $N=128^2$.
(a)--(c) Dependence on the contact energy $J_{00}$ at $h=h_{01}=h_{12}=h_{20}=1.3$.
The dynamic modes are described at the top of the figure.
(a) Probabilities of three states.
The gray up-pointing triangles or squares, blue down-pointing triangles or crosses, and red circles or diamonds represent the data for $s=0$, $1$, and $2$, respectively.
(b) Probabilities of one-phase state and three-phase coexistence.
The dashed lines in (a--b) represent kinetically trapped modes (SW+E$_1$ or E$_1$).
(c) Lifetimes of the homogeneous phases ($s=0$, $1$, and $2$) and SW mode.
(d) Dynamic phase diagram as a function of $J_{00}$ for various $h$.
The squares (light blue) and circles (gray) represent the HC and SW modes, respectively.
The up-pointing and down-pointing triangles (green) represent the homogeneous phases of $s=0$ and $s=2$ (E$_0$ and E$_2$), respectively.
The pluses (blue), open diamonds (magenta), closed diamonds (magenta), and crosses (red) represent the temporal coexistence:
SW+HC, SW+E$_0$, SW+E$_1$, SW+E$_2$, respectively.
The asterisks indicate the hysteresis of two modes (E$_0$ and SW+E$_1$ or E$_1$).}
\label{fig:j00}
\end{figure*}

Last, we consider a large system of $N=256^2$ sites, in which the HC mode only exists in the hysteresis at low $h$ in symmetric conditions.
The effects of the asymmetric flipping energies are similar to those in the small system ($N=128^2$).
As $h_{01}$ increases, the number ratio $N_2/N$ of the $s=2$ state increases in the SW mode, and subsequently, the E$_2$ phase appears via the temporal coexistence with the SW mode (SW+E$_2$),
as shown in Fig.~\ref{fig:n256}.
Since the size of biphasic domains and the number density of three-contact points continuously decrease with increasing $N_2/N$,
the transition between the SW and E$_\alpha$ modes likely occurs continuously via the mode coexistence at very large $N$.

\section{Dynamics for Asymmetric Contact Energies}\label{sec:asyme}

In this section, we describe the dynamics in the case of asymmetric contact energies $J_{00}\ne J_{11}=J_{22}=J$ 
and symmetric flipping energies $h_{01}=h_{12}=h_{20}=h$, for $N=128^2$.
When $J_{00}>J$, the $s=0$ state is stabilized and becomes dominant at high $J_{00}$, so that
the E$_0$ mode appears (Fig.~\ref{fig:j00}).
Conversely, with decreasing $J_{00}<J$, the E$_2$ mode appears since the nucleation of a $s=0$ domain is suppressed due to the low contact energy in the domain.
This is captured by the lifetimes of one-state dominant phases, as in the case of the asymmetric flipping energies (see Fig.~\ref{fig:j00}(c)).

\begin{figure*}[tbh]
\includegraphics[]{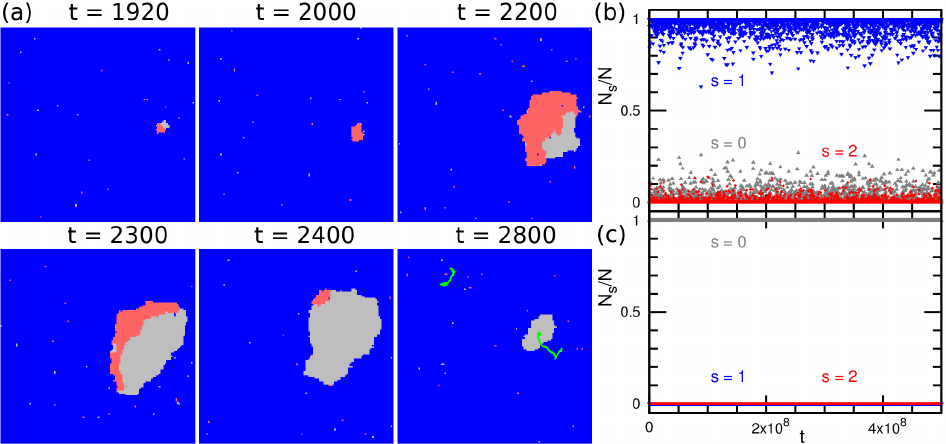}
\caption{
E$_1$ and E$_0$ modes in the hysteresis region at $J_{00}=2.5$,  $J_{11}=J_{22}=2$,  $h=h_{01}=h_{12}=h_{20}=1.3$, and $N=128^2$.
(a) Sequential snapshots in the E$_1$ mode.
A biphasic domain moves in the direction of the $s=2$ region.
The green lines in the last snapshot represent the trajectories of two biphasic domains consisting of $s=0$ and $2$ states. The right trajectory corresponds to the sequential snapshots in (a). The left trajectory corresponds to the domain shown in the last period of Movie S4.
(b) Time evolution of the fraction of each state (corresponding to snapshots in (a)).
The light gray, blue (dark gray), and red (medium gray) colors represent the data for $s=0$, $1$, and $2$, respectively.
(c) Time evolution of the fraction of each state starting from a $s=0$ dominant phase (E$_0$ mode).
}
\label{fig:tm13e25}
\end{figure*}

Interestingly, hysteresis occurs at high $h$ and high $J_{00}$. In the corresponding region of Fig.~\ref{fig:j00}(d) (see also Movie S4), either 
the E$_0$ mode or the E$_1$ (or SW+E$_1$) mode appears depending on the initial conditions.
Those modes do not change into each other in our simulation times $\sim 10^8$, since their lifetimes become longer due to high transition energy barriers.
In the E$_1$ mode, biphasic domains of $s=2$  and $s=0$ often appear but do not grow to cover the entire lattice 
(see Figs.~\ref{fig:tm13e25}(a) and (b)).
Interestingly, for $1.1\le h \le 1.3$, the SW+E$_1$ or E$_1$ modes, although they can be stabilized inside the hysteresis region, disappear in favor of the $E_0$ mode on \textit{both} sides of the hysteresis region (see Fig.~\ref{fig:j00}(b)).
Thus, the SW+E$_1$ mode is likely a kinetically trapped metastable state, while the E$_0$ mode seems to be the stable state.
In contrast, at $h\ge 1.5$,  
 the E$_0$ mode continues outside of the hysteresis region at higher $J_{00}$, while it is the SW+E$_1$ mode that continues on the other side at lower $J_{00}$,
see Fig.~\ref{fig:j00}(d).
The SW+E$_1$ mode becomes the E$_1$ mode in the middle of the hysteresis region ($J_{00}\ge 2.5$ at $h=1.5$).
Thus, the stable mode in the long-time limit likely changes from the SW+E$_1$ or E$_1$ mode to the E$_0$ mode in the middle of the hysteresis region.

In the symmetric condition ($J_{00}=J_{11}=J_{22}$ and  $h_{01}=h_{12}=h_{20}=h$),
three dominant phases ($s=0$, $1$, and $2$) are trapped in the limit $h \ll 1$.
In particular, they are degenerated thermal-equilibrium states at $h=0$.
The hysteresis found here is a nonequilibrium extension of this degeneration.

The ballistic motion of biphasic domains is clearly seen in Fig.~\ref{fig:tm13e25}(a) and Movie S4.
The trajectories of two domains are shown in the last snapshot of Fig.~\ref{fig:tm13e25}(a) (the right one corresponds to the sequential snapshots in  Fig.~\ref{fig:tm13e25}(a), and the left top one to the last period of Movie S4).
The right trajectory consists of two straight (ballistic) parts and a dangling (fluctuating at a location) part.
The first straight part corresponds to the motion of the small biphasic domain shown in the first snapshot of Fig.~\ref{fig:tm13e25}(a),
and the second corresponds to that of the large biphasic domain shown in the three middle snapshots of Fig.~\ref{fig:tm13e25}(a).
Therefore, a domain moves ballistically when it is biphasic, but it stops when it becomes a single phase.

\begin{figure}[tbh]
\includegraphics[]{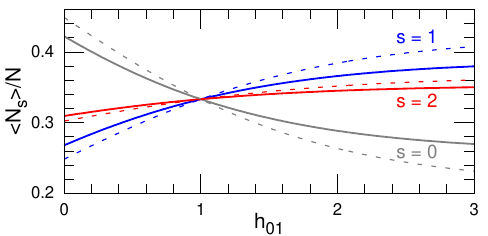}
\caption{
Density of the states $s=0$, $1$, and $2$ as a function of $h_{01}$ in an homogeneously mixed phase for $h_{12}=h_{20}=1$ and $J_{\alpha\alpha}=0$. Note that $h_{01}+h_{12}+h_{20}\ne0$, so the system is out of equilibrium. The solid and dashed lines represent the data for Metropolis rates and Glauber rates, respectively.
}
\label{fig:fix}
\end{figure}

\section{Phenomenological mean-field model}\label{sec:theory}

\subsection{Homogeneous Mixed State in the Absence of Interactions}\label{sec:structureless}

Let us call $\rho_\alpha=N_\alpha/N$ the number density of the state $s=\alpha$. Since there are no empty sites,
$\rho_0+\rho_1+\rho_2=1$.    
We work in units such that the thermal energy $k_{\mathrm{B}}T$ and the size of the lattice sites are unity.
Let us first consider a simplified situation in which we disregard all spatial structures and where the interactions between adjacent sites are neglected, i.e., $J_{\alpha\alpha}=0$.  In agreement with Eq.~(\ref{StartingHamiltonian}), the energy density (per site) is given by
\begin{equation}
f_{\mathrm{self}}=\rho_1\varepsilon_1+\rho_2\varepsilon_2+(1-\rho_1-\rho_2)\varepsilon_0,
\end{equation}
and the flipping energies $h_{\alpha\beta}$ are $\varepsilon_\alpha-\varepsilon_\beta$ in equilibrium.

As mentioned earlier, we place the system in an arbitrary nonequilibrium state, by setting $h_{\alpha\beta}=\varepsilon_\alpha-\varepsilon_\beta+h^{\mathrm{neq}}_{\alpha\beta}$, with $h_{\beta\alpha}=-h_{\alpha\beta}$, while assuming that the transition rates $w_{\alpha\beta}$ obey the \textit{local} detailed balance condition:
    $w_{\alpha\beta}/w_{\beta\alpha} =\exp(h_{\alpha\beta})$.
We therefore allow for $h_{01}+h_{12}+h_{20}=h_{\mathrm{cyc}}\ne0$, with $h_{\mathrm{cyc}}=h^{\mathrm{neq}}_{01}+h^{\mathrm{neq}}_{12}+h^{\mathrm{neq}}_{20}$, which we assume positive to favor the cycling $s=0\to1\to2\to0$. In practice, we take 
Metropolis rates
$w_{\alpha\beta}=\min[1,\exp(h_{\alpha\beta})]$ or Glauber rates $w_{\alpha\beta}=[1+\exp(-h_{\alpha\beta})]^{-1}$.

The dynamical equations are then
\begin{eqnarray}
\label{ro1}
    \dot\rho_1&=w_{01}\rho_0-\left(w_{10}+w_{12}\right)\rho_1+w_{21}\rho_2,\\
    \label{ro2}
    \dot\rho_2&=w_{02}\rho_0+w_{12}\rho_1-\left(w_{20}+w_{21}\right)\rho_2,\\
    \dot\rho_0&=-\left(w_{01}+w_{02}\right)\rho_0+w_{10}\rho_1+w_{20}\rho_2.
\end{eqnarray}
The third equation follows from the first two due to the density conservation equation. In the stationary state, $\dot\rho_\alpha=0$, hence replacing $\rho_0$ by $1-\rho_1-\rho_2$ in Eqs.(\ref{ro1})--(\ref{ro2}) yields a linear system for the stationary densities $\rho_1$ and $\rho_2$, which gives
\begin{eqnarray}
    \frac{\rho_1}{\rho_0} &=\frac
    {w_{02}w_{21}+w_{01}(w_{20}+w_{21})}
    {w_{12}w_{20}+w_{10}(w_{20}+w_{21})}, \\
    \frac{\rho_2}{\rho_0} &=\frac
    {w_{01}w_{12}+w_{02}(w_{10}+w_{12})}
    {w_{12}w_{20}+w_{10}(w_{20}+w_{21})}.
    \label{stat2}
\end{eqnarray}
For Metropolis rates, assuming $h_{01}\ge 0$, $h_{12}\ge 0$, and $h_{20}\ge 0$, we obtain
\begin{eqnarray}
\label{statMetro}
    \frac{\rho_1}{\rho_0}
    &=e^{h_{01}}
    \frac
    {1+e^{h_{12}}+e^{h_{02}}}
    {1+e^{h_{12}}+e^{h_{02}}e^{h_{\mathrm{cyc}}}}, \\
    \frac{\rho_0}{\rho_2} &=e^{h_{20}}
    \frac
    {1+e^{h_{01}}+e^{h_{21}}}
    {1+e^{h_{01}}+e^{h_{21}}e^{h_{\mathrm{cyc}}}}.
\end{eqnarray}
Equilibrium is achieved for $h_{\mathrm{cyc}}=0$. In this case, Eq.~(\ref{statMetro})  give $\rho_1/\rho_0=\exp(h_{01})$ and $\rho_0/\rho_2=\exp(h_{02})$, which produces the Boltzmann distribution $\rho_\alpha\propto\exp(-\varepsilon_\alpha)$, as expected.

When $h_{\mathrm{cyc}}\ne0$, the stationary densities differ from the equilibrium ones.
As $h_{01}$ increases at fixed $h_{12}$ and $h_{20}$, we find that $\rho_1$ increases and $\rho_0$ decreases, as expected (see Fig.~\ref{fig:fix}).
When  Glauber rates are used, the density changes become slightly larger (compare the dashed and solid lines in Fig.~\ref{fig:fix}).

\subsubsection{Equilibrium State and Free Energy}
\label{sec:EquilibriumState}

At equilibrium, i.e., for $h_{\mathrm{cyc}}=0$, the densities can be obtained from the  free-energy density $f=f_{\mathrm{self}}+f_{\mathrm{mix}}$, with
\begin{equation}
f_{\mathrm{mix}}=\rho_1\ln\rho1+\rho_2\ln\rho_2+(1-\rho_1-\rho_2)\ln(1-\rho_1-\rho_2),
\end{equation}
arising from the entropy of mixing.
Indeed, minimizing $f$ with respect to $\rho_1$ and $\rho_2$ yields  $\rho_\alpha/\rho_0=\exp(\varepsilon_0-\varepsilon_\alpha)$.

Note that while the equilibrium densities are obtained by minimizing the full free energy density $f$, the transition rates $w_{\alpha\beta}$ result from the variations of $f_{\mathrm{self}}=f-f_{\mathrm{mix}}$.

\subsection{Continuum Theory with Spatial Gradients and Interactions}

We now assume that the densities $\rho_i(\mathbf{x})$ vary in space. Since we consider only local state flips (no state exchange between adjacent sites), the dynamical equations still have the same form locally:
\begin{eqnarray}
\label{rho1dot}
    \dot\rho_1(\mathbf{x})&=w_{01}\rho_0-\left(w_{10}+w_{12}\right)\rho_1+w_{21}\rho_2,\\
    \dot\rho_2(\mathbf{x})&=w_{02}\rho_0+w_{12}\rho_1-\left(w_{20}+w_{21}\right)\rho_2,
    \\
    \rho_0(\mathbf{x})&=1-\rho_1-\rho_2.
    \label{rho0dot}
\end{eqnarray}
The transition rates $w_{ij}(\mathbf{x})$, however, now depend on the local free-energy variations, including the interactions between adjacent sites and the penalty due to density gradients. The latter are required in order to take into account that the nucleation of a new phase behaves differently within a domain and at the interface between two domains. Thus, the free-energy density is now $f=f_{\mathrm{self}}+f_{\mathrm{mix}}+f_{\mathrm{int}}+f_{\mathrm{grad}}$, with
\begin{eqnarray}
    f_{\mathrm{int}}&=&-\frac{u_1}2\rho_1^2-\frac{u_2}2\rho_2^2-\frac{u_0}2\left(1-\rho_1-\rho_2\right)^2,\\
    f_{\mathrm{grad}}&=&
    \frac k2\left(\nabla\rho_1\right)^2+
    \frac k2\left(\nabla\rho_2\right)^2+
    \frac k2\left[\nabla\left(1-\rho_1-\rho_2\right)\right]^2, \hspace{0.4cm}
\end{eqnarray}
where $u_\alpha>0$ quantifies the attractive couplings between like states, and $k>0$ the penalty associated with density gradients (common to all states for simplicity).

\begin{figure*}
\includegraphics[width=15.5cm]{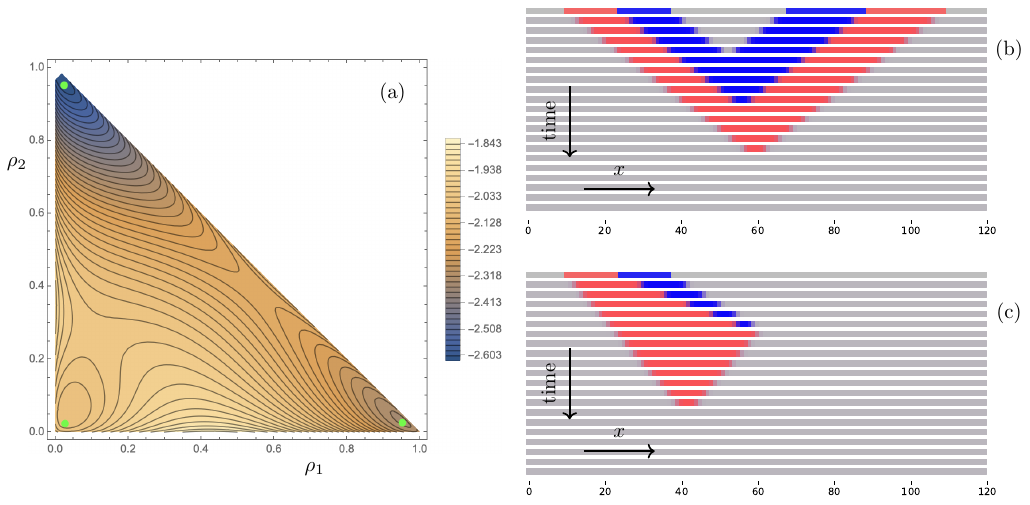}
\caption{
Dynamics in the continuous model.
(a)-(b) Symmetric case $h_{01}=h_{12}=h_{20}=0.3$, obtained from $\varepsilon_0=0$, $\varepsilon_1=-0.3$, $\varepsilon_2=-0.6$, and $h_{20}^{\mathrm{neq}}=h_{\mathrm{cyc}}=0.9$. The interaction and gradient coefficients are $u=4$ and $k=1$, respectively. (a) Free energy landscape $f$ as a function of $\rho_1$ and $\rho_2$. The bottom left, bottom right and top left minima correspond to the phases $s=0$, $s=1$, and $s=2$, respectively. (b) Numerical integration of Eqs.~(\ref{rho1dot})--(\ref{rho0dot}) in one dimension, showing two different-sized $s=2,1$ (red, blue) biphasic solitary waves moving in opposite directions, colliding and annihilating each other. Time flows from top to bottom. The time lapse between two snapshots: $4\times10^5$ time steps $dt=1.25\times10^{-4}$. The RGB color encodes the system state as $\rho_0\mathcal{G}+\rho_1\mathcal{B}+\rho_2\mathcal{R}$, with $\mathcal{G}$ gray, $\mathcal{B}$ blue, and $\mathcal{R}$ red.
(c) Asymmetric case. Evolution of a biphasic $s=1,2$ band in a $s=0$ backgroud for asymmetric flipping energies: $h_{01}=0.4$, $h_{12}=0.45$, and $h_{20}=0.3$, obtained from $\varepsilon_0=0$, $\varepsilon_1=-0.4$, $\varepsilon_2=-0.85$, and $h_{20}^{\mathrm{neq}}=h_{\mathrm{cyc}}=1.15$. As the band propagates, the blue phase ($s=1$) recedes, and after its disappearance, the red phase ($s=2$) is in turn consumed by the grey phase ($s=0$).
}
\label{fig:cmod}
\end{figure*}

As discussed in Sec.~\ref{sec:EquilibriumState}, the transition rates must be calculated from the free energy minus the contribution arising from the entropy of mixing. The latter is given by the functional
\begin{equation}
F[\rho_1,\rho_2]=\int\!d^2x\left(f_{\mathrm{self}}+f_{\mathrm{int}}+f_{\mathrm{grad}}\right).
\end{equation}
The  energy change associated with the flip of one site from the state $s=0\to1$, at point $\mathbf{x}$, supplemented by the nonequilibrium contribution $h_{10}^{\mathrm{neq}}$, is given by
\begin{eqnarray}
    \Delta H_{01}(\mathbf{x}) &=& \left.\frac{\delta F}{\delta \rho_1(\mathbf{x})}\right|_{\rho_2}   \nonumber \\
 &=&\varepsilon_1-\varepsilon_0-u_1\rho_1+u_0\rho_0 
   -k\nabla^2\left(\rho_1-\rho_0\right)-h_{01}^{\mathrm{neq}}   \nonumber \\ 
 &=& -\Delta H_{10}(\mathbf{x}). \label{h10eq}
\end{eqnarray}
Likewise, $\Delta H_{02}=\varepsilon_2-\varepsilon_0-u_2\rho_2+u_0\rho_0-k\nabla^2(\rho_2-\rho_0)-h_{02}^{\mathrm{neq}}=-\Delta H_{20}$. From the energy change in the sequence $s=2\to0\to1$, we infer also $\Delta H_{21}=\varepsilon_1-\varepsilon_2-u_1\rho_1+u_2\rho_2-k\nabla^2(\rho_1-\rho_2)-h_{21}^{\mathrm{neq}}=-\Delta H_{12}$.

In the following, we shall take $u_0=u_1=u_2=u$ and break detailed balance by setting $h_{01}^{\mathrm{neq}}=0$, $h_{12}^{\mathrm{neq}}=0$, and $h_{20}^{\mathrm{neq}}=h_{\mathrm{cyc}}>0$. For convenience, we assume Glauber rates:
\begin{equation}
\label{localglauber}
    w_{\alpha\beta}(\mathbf{x})=\frac1{1+e^{\Delta H_{\alpha\beta}(\mathbf{x})}}.
\end{equation}
Note that they obey the local detailed balance condition $w_{\alpha\beta}/w_{\beta\alpha}=\exp(-\Delta H_{\alpha\beta})$, even in the nonequilibrium case $h_{\mathrm{cyc}}\ne0$.

\subsubsection{Numerical Analysis}

Equations~(\ref{rho1dot})--(\ref{rho0dot}) with the rates of Eq.~(\ref{localglauber}), obtained from the energy variations $\Delta H_{\alpha\beta}(\mathbf{x})$ defined in Eq.~(\ref{h10eq}) and below, allow to calculate the spatially resolved evolution of the system. We have integrated the discretized version of these equations in one dimension with periodic boundary conditions in a lattice of $L=120$ sites using the classical Runge-Kutta method. The first and second derivatives were discretized using central differences with fourth-order accuracy. 

We first investigated the dynamics in cyclic-symmetry conditions, with $h_{01}=h_{12}=h_{20}=0.3$, in which case the three states are equivalent and the dynamics promotes the sequence $s=0\to1\to2\to0$. A biphasic $s=1,2$ band in a $s=0$ background forms a travelling wave, which behaves as a perfect soliton~\cite{nogu24}.  The stationary densities $(\rho_1,\rho_2)$ of the three phases are indicated as the green points in Fig.~\ref{fig:cmod}(a), matching the minima of the free energy $f$. Such travelling waves stochastically appear also in two dimensions, as can be seen in Fig.~\ref{fig:e12a} and Movie S1. They also produce the spiralling dynamics around the contact point of three different domains (see Figs.~\ref{fig:e12a} and \ref{fig:n256}(d)). 

Biphasic bands travelling in opposite directions annhilate each other, as shown in Fig.~\ref{fig:cmod}(b), even when their sizes are different, as a natural consequence of the $s=0\to1\to2\to0$ transformation. Such events can also be seen in two dimensions (see Fig.~\ref{fig:n256}(d) and Movie S1).

In the case of asymmetric flipping energies, we find that three interfaces $s=0\to 1$, $s=1\to 2$ and $s=2\to 0$ have different velocities. Depending on the relative asymmetries, this can lead to the widening of a biphasic band or its disappearance (see Fig.~\ref{fig:cmod}(c)).

\section{Summary}\label{sec:sum}

We have studied the dynamics of the active cyclic Potts model under asymmetric conditions.
We found that biphasic domains and non-cyclic one-state dominant phases are induced by
either asymmetric flipping energies or asymmetric contact energies.
The spiral wave mode can temporally coexist with either the one-state dominant phases or the homogeneous cycling mode.
Biphasic domains move like amoeba and exhibit division and disappearance.
When the flipping energy from the $s=0$ to the $s=1$ state is increased, or the contact energy between $s=0$ sites is decreased 
while other flipping energies are fixed,
the $s=2$ dominant phase appears owing to the suppression of the nucleation of $s=0$ domains in the $s=2$ phase.
Two separate modes can be formed depending on the initial state due to hysteresis: one-state dominant phase and coexistence of spiral waves and one-state dominant phase, or two types of one-state dominant phases.

In reaction-diffusion systems, the length scales of spatiotemporal patterns are usually controlled by the diffusion coefficients of the reactants~\cite{mikh09}.
However, in the present system, the diffusion of the various states is not taken into account.
The observed wavelengths fluctuate largely and are determined, on average, by the competition between domain nucleation and growth.
We mentioned three experimental situations relevant to the present cyclic model in the Introduction, including reaction on a catalytic surface.
Although oxidation or reduction on catalytic surfaces has been intensively studied in experiments~\cite{ertl08,bar94,goro94,barr20,zein22}, the effects of nucleation and growth have not been well understood.
We believe that the rich dynamics of the present system, including the amoeba-like motion of biphasic domains, can be experimentally observed on catalytic surfaces.
Here, we simulated small systems.
Such small systems can be realized using metal nanoparticles~\cite{tang20}.
Chemical waves are known to induce shape deformations of metal particles~\cite{tang20,ghos22}, lipid membranes~\cite{wu18,tame21,nogu23a}, and macroscopic gel sheets~\cite{lono14,maed08,levi20}.
In this study, we have only considered non-deformable flat surfaces. Surface deformation and the dynamics on curved surfaces are likely to modify the dynamics of the active cyclic Potts model and yield new phenomena.

Very recently, Manacorda and Fodor proposed a different type of thee-state lattice model~\cite{mana23}.
Their model allows the overlap of particles in each site
and particles interact only with the other particles in the same site.
The correlation of the neighbor sites is caused only by particle diffusion.
They reported spiral wave and oscillatory modes but not homogeneous cycling mode.
However, in the parameter range they call the arrest phase (one state covers the entire lattice), 
the homogeneous cycling might appear in longer simulations of their model.

Here, we used the three-state Potts model in the square lattice.
It is applicable systematically from thermal equilibrium to far-from-equilibrium.
The dynamics may be changed when other types of lattices are used.
When four or more states are considered,
more complex dynamics are expected.
Hence, the cyclic Potts model is an excellent model system to study nonequilibrium dynamics coupled with nucleation and growth.
It is simple and easy to extend to several directions for further studies.

\begin{acknowledgments}
We thank Fr\'ed\'eric van Wijland
for stimulating discussion.
This work was supported by JSPS KAKENHI Grant Number JP21K03481 and JP24K06973. 
\end{acknowledgments}


%

\end{document}